\documentclass[conference]{IEEEtran}
\IEEEoverridecommandlockouts

\usepackage{cite}
\usepackage{amsmath,amssymb,amsfonts}
\usepackage{algorithmic}
\usepackage{graphicx}
\usepackage{textcomp}
\usepackage{xcolor}
\usepackage{hyperref}
\usepackage{booktabs}
\usepackage{multirow}
\usepackage{listings}
\def\BibTeX{{\rm B\kern-.05em{\sc i\kern-.025em b}\kern-.08em
    T\kern-.1667em\lower.7ex\hbox{E}\kern-.125emX}}

\begin{document}

\title{Toward Human-Centered Interactive Clinical Question Answering System}

\author{\IEEEauthorblockN{Dina Albassam}
\IEEEauthorblockA{\textit{Department of Computer Science} \\
\textit{University of Illinois Urbana-Champaign} \\
Email: dinasa2@illinois.edu}}

\maketitle

\begin{abstract}
Unstructured clinical notes contain essential patient information but are challenging for physicians to search and interpret efficiently. Although large language models (LLMs) have shown promise in question answering (QA), most existing systems lack transparency, usability, and alignment with clinical workflows. This work introduces an interactive QA system that enables physicians to query clinical notes via text or voice and receive extractive answers highlighted directly in the note for traceability.

The system was built using OpenAI models with zero-shot prompting and evaluated across multiple metrics, including exact string match, word overlap, SentenceTransformer similarity, and BERTScore. Results show that while exact match scores ranged from 47\% to 62\%, semantic similarity scores exceeded 87\%, indicating strong contextual alignment even when wording varied.

To assess usability, the system was also evaluated using simulated clinical personas. Seven diverse physician and nurse personas interacted with the system across scenario-based tasks and provided structured feedback. The evaluations highlighted strengths in intuitive design and answer accessibility, alongside opportunities for enhancing explanation clarity.
\end{abstract}

\begin{IEEEkeywords}
clinical QA, medical NLP, user-centered design, LLMs, electronic health records
\end{IEEEkeywords}

\section{Introduction}

Electronic health records (EHRs) have become foundational to modern healthcare, serving as digital repositories of patient information. By 2015, over 96\% of non-federal acute care hospitals in the United States had adopted certified EHR systems to manage clinical data. These systems offer valuable opportunities for both individual patient care and broader data-driven insights. While structured components such as lab results and demographics are useful, the majority of clinically rich information resides in unstructured formats—namely, clinical notes written by physicians, including progress summaries, discharge instructions, and diagnostic impressions.

Unstructured clinical text captures nuanced reasoning, contextual decisions, and temporal sequences that are essential to understanding patient status. This makes it an ideal target for large language models (LLMs), which excel at processing free-text inputs~\cite{b5}. As a result, medical question answering (QA) systems powered by LLMs have emerged as promising tools to help clinicians extract relevant insights from complex patient records. For example, a system that can answer, ``Has the patient ever taken Atropine for hypertension?'', by reviewing multiple clinical notes, could significantly reduce the cognitive burden of manual chart review and support informed clinical decisions.

Despite this promise, current medical QA systems often fall short in practice. Clinicians report difficulties using these tools effectively—especially when interacting with unstructured notes—due to issues related to usability, interpretability, and trust. As Kell et al.~\cite{b13} highlighted in a systematic review, only 7 out of 79 biomedical QA systems had undergone any user evaluation, and none incorporated essential features like uncertainty estimation or source attribution. Most systems remain confined to technical research prototypes, lacking integration into clinical workflows and failing to meet the practical needs of healthcare professionals.

Existing QA research has explored a wide range of data types and approaches. Structured QA systems such as quEHRy~\cite{b4} and DrugEHRQA~\cite{b2} convert clinician queries into database-friendly formats, enabling high precision but limited flexibility. In the domain of unstructured data, Elgedawy et al.~\cite{b3} and Yu et al.~\cite{b5} introduced retrieval-augmented generation (RAG) methods to answer questions based on clinical notes. However, these systems offer little in terms of interactive design or physician-centered evaluation. Similarly, Jin et al.~\cite{b4} used templated QA techniques for specific extraction tasks, offering limited adaptability.

Outside clinical notes, systems such as MedQA~\cite{b6}, AskHERMES~\cite{b9}, and SemBioNLQA~\cite{b10} focus on medical literature, and while they have demonstrated value, they do not address the challenge of navigating free-text clinical narratives. Patient-oriented tools like PaniniQA~\cite{b11} and README~\cite{b12} aim to simplify discharge summaries for lay comprehension but are not designed for physician use.

Altogether, these efforts underscore a persistent gap: few QA systems provide end-to-end, interactive support for physicians querying free-text clinical notes. Furthermore, existing evaluation strategies do not adequately reflect real-world clinical utility.

To address these challenges, this study explores the following research questions:
\begin{itemize}
    \item How can interactive clinical QA systems be designed to support physicians’ information-seeking behaviors, particularly in enhancing the trustworthiness and transparency of model-generated answers?
    \item What are physicians’ perceptions of usability, interpretability, and trust when interacting with QA tools in clinical workflows?
    \item How do simulated agent-based (AI persona) evaluations compare to traditional human user studies in assessing QA systems’ effectiveness and utility?
\end{itemize}

In response to these questions, this study contributes the following:
\begin{itemize}
    \item A comparative evaluation of OpenAI models using zero-shot prompting on the emrQA-msquad dataset to select the most suitable model for clinical QA.
    \item The design and implementation of an interactive QA system that allows physicians to ask questions via text or voice and highlights exact answer spans in clinical notes to promote transparency and trust.
    \item An initial user evaluation using AI personas, which simulate real clinical scenarios to assess the system’s usability and effectiveness in a scalable, automated manner.
\end{itemize}

This work aims to bridge the gap between technical QA advancements and real-world clinical applicability, laying the foundation for trustworthy, physician-centered QA systems for unstructured medical notes.

\section{Related Work}

\subsection{Clinical QA Systems}

Medical QA systems have increasingly leveraged large language models (LLMs) to reason over both structured and unstructured electronic health record (EHR) data. Structured QA systems such as quEHRy~\cite{b1} and DrugEHRQA~\cite{b2} convert clinician queries into formal database queries using concept normalization and SQL translation, achieving high precision. DrugEHRQA also introduces a benchmark dataset combining structured and unstructured sources to support multimodal QA development.

In contrast, unstructured QA systems use retrieval-augmented generation (RAG) techniques to process narrative clinical data. Elgedawy et al.~\cite{b3} and Yu et al.~\cite{b5} proposed RAG pipelines—one as a chatbot interface and another using a federated architecture (FedRAG)—to provide contextual answers from clinical notes. Jin et al.~\cite{b4} implemented a templated QA approach to extract information about injection drug use, demonstrating the utility of QA for specific clinical tasks.

Other systems focus on broader medical knowledge. MedQA~\cite{b6} and AskHERMES~\cite{b9} answer definitional queries using MEDLINE, showing improved clinician satisfaction over general-purpose tools like Google. SemBioNLQA~\cite{b10} supports multiple question types (yes/no, factoid, list, summary) and performs competitively in BioASQ using semantic retrieval and UMLS-based techniques.

Some QA tools are patient-centric. PaniniQA~\cite{b11} enhances patient understanding of discharge summaries through quizzes, and README~\cite{b12} generates simplified, contextual lay definitions. Both systems are evaluated with patients and demonstrate the impact of QA on comprehension.

Despite these innovations, most systems remain technical prototypes. A review by Kell et al.~\cite{b13} found only 7 of 79 biomedical QA systems underwent any user evaluation, and none supported features like uncertainty estimation or source attribution. This underscores a critical gap in usability, interpretability, and trustworthiness—leaving physician-centered QA systems underdeveloped.

\subsection{Information-Seeking Needs of Physicians}

Effective QA system design must consider how clinicians search and interpret information during care. Vong and Then~\cite{b15} assessed the CliniCluster system using the PICO framework and found that clinicians often ask vague or evolving questions and benefit from tools supporting query refinement and semantic exploration. AskHERMES~\cite{b9} and MedQA~\cite{b6} similarly highlight that clinicians value clear, fast, and context-aware answers. These insights point to the importance of aligning QA system design with physicians’ cognitive workflows—not just technical accuracy.

\subsection{User Evaluation Approaches}

Although most studies prioritize benchmark performance, few assess real-world utility. AskHERMES~\cite{b9} and MedQA~\cite{b6} incorporated cognitive evaluations comparing their tools to Google and PubMed, but such evaluations remain rare. Elgedawy et al.~\cite{b3} developed an interactive QA system but did not conduct user studies. More recently, IQA-EVAL~\cite{b16} proposed using simulated ``AP personas'' to evaluate interactive QA systems. These personas simulate user behavior across tasks and offer a scalable, repeatable way to assess usability and relevance—providing a practical alternative to resource-intensive human evaluations.

\section{Methods}

\subsection{Problem Formulation}
The interactive clinical QA task is framed as an extractive question answering problem over unstructured clinical notes. Given a natural language question posed by a physician and a corresponding clinical note, the system is expected to retrieve a verbatim answer span from within the note. For example, when presented with the question ``What medications was the patient prescribed at discharge?'' and a note stating:

\textit{Pt is a 65 y/o male with a history of CHF and hypertension. Discharged on Lisinopril 10mg daily and Metoprolol 25mg BID...}

the system should extract: \textit{``Lisinopril 10mg daily and Metoprolol 25mg BID''}.

Beyond answer retrieval, the system supports:
\begin{itemize}
    \item Multi-modal input (typed or spoken queries),
    \item Highlighting of answer spans in the original note,
    \item Handling of vague or evolving queries through suggested prompts
\end{itemize}

\subsection{Dataset}
The emrQA-msquad dataset~\cite{b11} was used to train and evaluate the QA model. This dataset fuses the structural rigor of SQuAD v2.0 with domain-specific clinical content from emrQA, providing 163{,}695 questions aligned with 4{,}136 manually curated answers from EMRs. Its span-based answer format makes it well-suited for extractive QA over clinical notes. In contrast, alternative datasets like MedQuAD and MedQA, with multiple-choice structures, are less compatible with extractive methodologies. 

For our evaluation, we used a randomly sampled subset of \textbf{4,000 rows} from emrQA-msquad to ensure manageable runtime and efficient comparison across models.

\subsection{System Overview}
\begin{figure}[!t]
\centerline{\includegraphics[width=\linewidth]{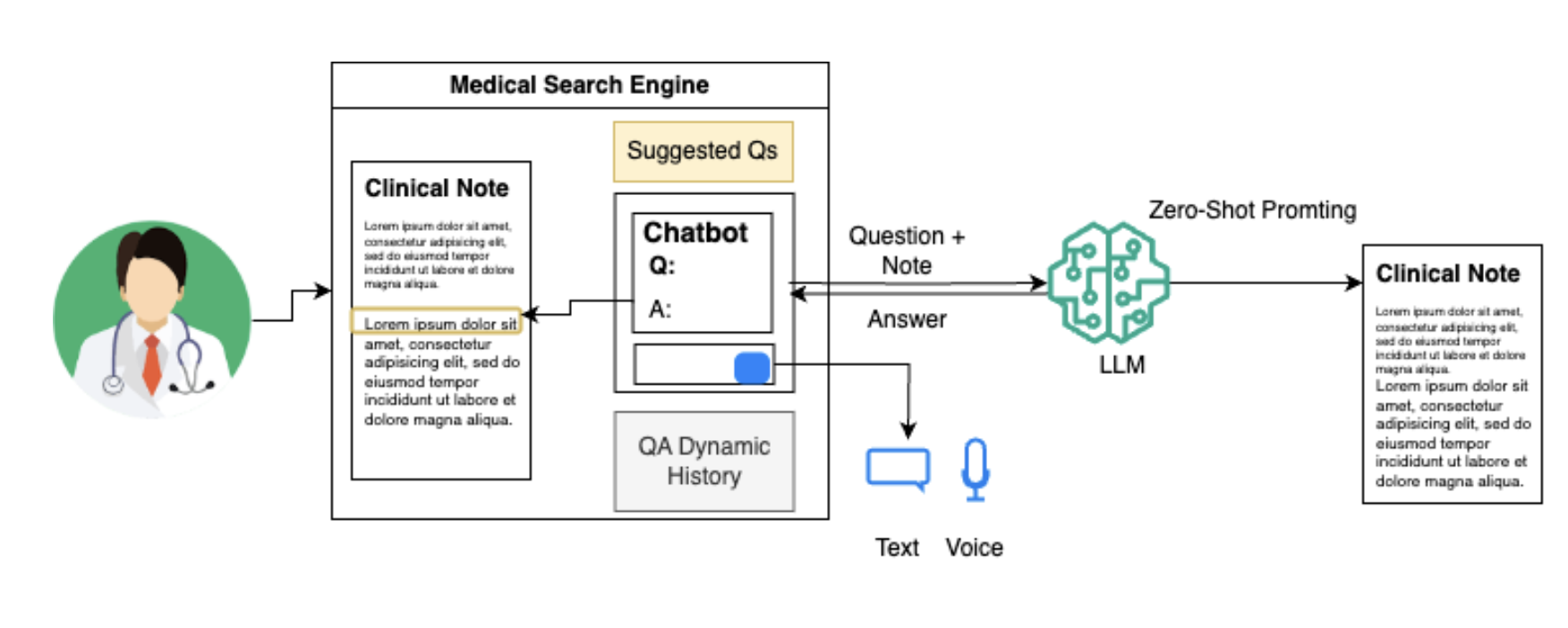}}
\caption{Overview of the Medical QA system. Questions are submitted by the user, paired with the selected note, and passed to the LLM. The extracted answer is highlighted for transparency.}
\label{fig:qa-architecture}
\end{figure}

The QA system consists of a web-based interface backed by a lightweight Flask server. As illustrated in Fig.~\ref{fig:qa-architecture}, clinicians can explore clinical notes, pose natural language questions via text or voice, and receive extractive answers. The model response is highlighted directly within the clinical note to support interpretability.

\subsection{Modeling}

\subsubsection{Zero-Shot Prompting}
To evaluate baseline model performance, a zero-shot prompting approach was employed using OpenAI models hosted on Azure. The prompt format explicitly instructs the model to extract a verbatim span from the context without generating or paraphrasing, ensuring alignment with extractive QA requirements. A deterministic setup was enforced using a low temperature (0.1). Batch processing and retry logic were integrated to maintain throughput and robustness during large-scale evaluation.

The exact prompt used was structured as follows:

\begin{lstlisting}
{"role": "system", "content": "You are an AI assistant that helps doctors answer clinical questions based on patient contexts."}
{"role": "user", "content": "Extract the exact answer span from the provided context that directly answers the question. The answer must be a contiguous span of text copied verbatim from the context. Do not generate or infer. Do not include any explanation."}
\end{lstlisting}

\subsubsection{Performance Evaluation}
Model outputs were compared against gold-standard spans using both lexical and semantic metrics:

\begin{itemize}
    \item \textbf{Exact String Similarity:} Uses Python’s \texttt{SequenceMatcher} to calculate character-level match precision.
    \item \textbf{Word Overlap:} Computes token overlap between the predicted and reference spans.
    \item \textbf{Normalized Word Overlap:} Normalizes both spans (lowercasing, punctuation removal) before calculating overlap.
    \item \textbf{Sentence Similarity:} Applies cosine similarity using \texttt{SentenceTransformer} (all-MiniLM-L6-v2).
    \item \textbf{BERTScore:} Evaluates contextual token-level alignment using BERT embeddings.
\end{itemize}

\subsubsection{Evaluation Results}
Three OpenAI models were evaluated on a held-out set from emrQA-msquad. Table~\ref{tab:results} presents their comparative performance across metrics.

\begin{table}[!h]
\caption{Model Performance on emrQA-msquad Dataset}
\label{tab:results}
\centering
\begin{tabular}{|l|c|c|c|c|c|}
\hline
\textbf{Model} & \textbf{Exact} & \textbf{Word} & \textbf{Norm.} & \textbf{ST Sim.} & \textbf{BERTScore} \\
\textbf{} & \textbf{Match} & \textbf{Overlap} & \textbf{Overlap} & (\%) & (\%) \\
\hline
OpenAI o3-mini & 62.44 & 49.96 & 49.96 & 70.13 & 90.70 \\
GPT-4o         & 59.05 & 46.40 & 46.40 & 68.86 & 89.64 \\
gpt-35-turbo-16k & 47.00 & 39.61 & 39.61 & 60.62 & 87.89 \\
\hline
\end{tabular}
\end{table}

The results demonstrate that while exact match scores remain moderate, semantic alignment is strong—suggesting that the model often captures the intended meaning even when the phrasing differs from the reference answer. Among the evaluated models, \textbf{OpenAI o3-mini outperformed all others across every metric}, achieving the highest BERTScore of 90.70\%. This strong performance—particularly in semantic similarity—motivated our choice of \textbf{o3-mini} for system development.

\subsection{System Architecture and Development}

\begin{figure}[!h]
\centering
\includegraphics[width=\linewidth]{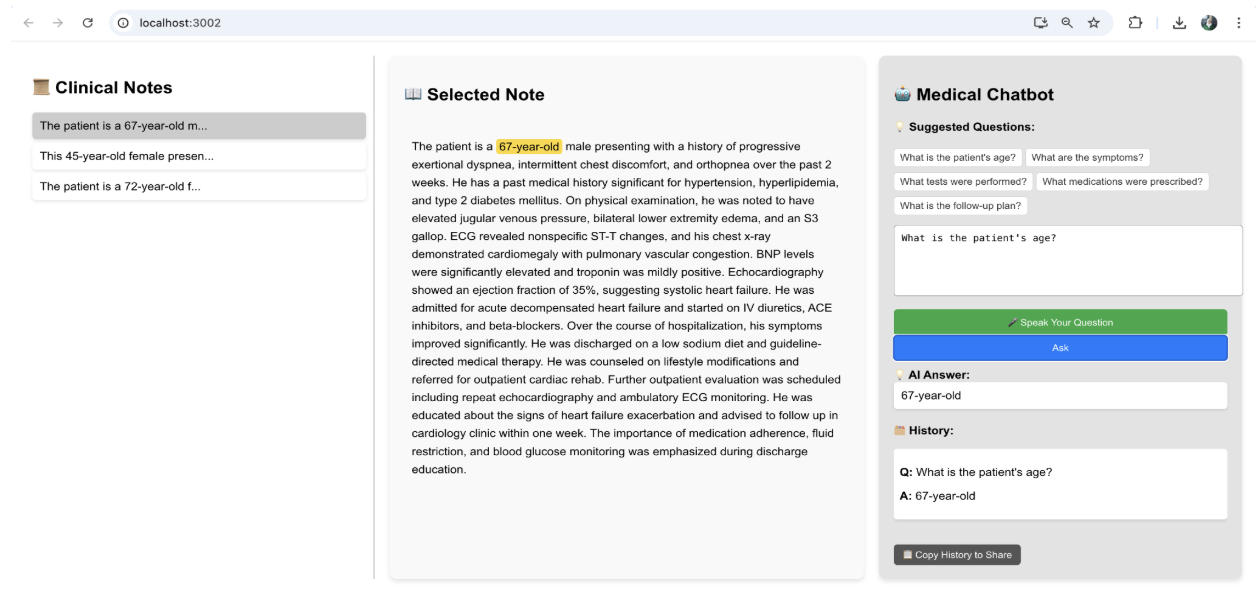}
\caption{Screenshot of the Clinical QA system interface. Users can select a patient note (left), view the full note with highlighted answers (center), and ask questions via text or speech through the chatbot (right).}
\label{fig:ui-screenshot}
\end{figure}

To support physician-centered question answering over clinical notes, an interactive Medical QA system was developed, consisting of a web-based frontend and a lightweight backend that communicates with a large language model (LLM) using zero-shot prompting. The system enables clinicians to explore patient notes, ask natural language questions using voice or text, and receive extractive, traceable answers highlighted directly in the note. A visual overview of the interface is provided in Figure~\ref{fig:ui-screenshot}, showing the note selection panel, answer highlighting, and chatbot interaction area.

\subsubsection{Backend: Flask-Based QA and Retrieval Services}
The backend is implemented using Flask and exposes several RESTful endpoints to support core functionalities such as clinical note retrieval, question answering, and suggested prompt generation. Clinical notes—containing patient histories, diagnoses, and medications—are stored in a JSON file and loaded at runtime to provide context for QA operations.

The system uses a fixed structured prompt format that is dynamically populated with the user’s question and the selected clinical note. This prompt is then sent to a language model for extractive answer generation. Retry logic is included to ensure robustness against transient API failures.

Endpoints are available to retrieve all notes or a specific note by ID. A dedicated QA endpoint dynamically combines the user’s question and note content for inference, while another endpoint provides suggested questions relevant to the selected note

\subsubsection{Frontend: React-Based Clinical QA Interface}
The user interface, built with React.js, presents an intuitive, three-panel layout:
\begin{itemize}
    \item \textbf{Left panel:} Lists clinical notes for selection.
    \item \textbf{Center panel:} Displays the selected note, with answers highlighted directly within the note.
    \item \textbf{Right panel:} Hosts the chatbot interface, allowing typed or spoken input. Suggested questions update dynamically based on the note.
\end{itemize}
The system displays the model's answer below each query and logs the QA interaction history. Users can export this history for further use.

\section{User Evaluation with AI-Personas}
To assess the usability and clinical utility of the MedicalQA system in a simulated real-world setting, an automated user evaluation was conducted using AI-generated personas. These personas were designed to reflect a diverse set of healthcare professionals varying in roles, specialties, experience levels, and technological comfort.

\subsection{Persona Design and Demographics}
Seven AI personas were created, including five physicians and two nurses. These personas spanned a range of clinical departments such as internal medicine, emergency, psychiatry, and intensive care, and represented different career stages—from junior practitioners to senior clinicians. Each persona had unique information-seeking behaviors and preferences shaped by age, experience, and clinical environment.

\begin{itemize}
    \item \textbf{Role distribution:} 5 physicians, 2 nurses
    \item \textbf{Age and experience variation:} Seniors (~50s), mid-level (~40s), juniors (~30s)
    \item \textbf{Tech comfort:} From low to high based on prior EHR and AI exposure
\end{itemize}

\subsection{Evaluation Setup}
Each AI persona was assigned a specific clinical scenario and prompted to interact with the MedicalQA system to retrieve answers from clinical notes. To simulate realistic use, the prompt included a screenshot of the system's user interface, allowing the model to evaluate the visual and functional aspects of the interaction. GPT-4o was instructed to embody each persona’s behavior and assess the system using three dimensions:
\begin{itemize}
    \item \textbf{Usability:} Are the questions and interface features intuitive?
    \item \textbf{Efficiency:} How quickly and effectively can the system retrieve relevant answers?
    \item \textbf{Feedback:} How helpful and clear is the presentation of answers?
\end{itemize}

\subsection{Persona Results}
Each simulated persona task yielded structured feedback. For instance, a primary care physician was able to identify the patient’s age, symptoms, and medications efficiently, reporting that question suggestions matched the task well and answer clarity was helpful. More experienced personas valued simplicity, while junior users prioritized explainability.

\subsection{Aggregate Evaluation Insights}
Persona feedback surfaced key system strengths and areas for improvement:
\begin{itemize}
    \item \textbf{Usability:} High for direct queries; moderate for complex or detailed extractions
    \item \textbf{Efficiency:} Strong for simple data; slower for unstructured inputs
    \item \textbf{Trust and Interpretability:} Accurate for facts but limited in diagnostic reasoning or explanation
    \item \textbf{Design Suggestions:} Add highlights, urgency flags, explainable outputs
\end{itemize}

\begin{table}[!h]
\caption{Aggregate Persona Evaluation Scores (out of 5)}
\label{tab:persona-scores}
\centering
\begin{tabular}{|l|c|}
\hline
\textbf{Metric} & \textbf{Avg. Score} \\
\hline
Usability & 4.0 \\
Efficiency & 3.9 \\
Trust / Interpretability & 3.4 \\
\hline
\end{tabular}
\end{table}

\section{Discussion and Limitations}
This work presents the development of an interactive QA system grounded in clinical notes, with evaluation via simulated AI personas. The system offers novel usability enhancements but also exposes several key limitations.

\subsection{Modeling Evaluation Challenges}
Selecting appropriate evaluation metrics remains complex. While semantic measures like BERTScore and SentenceTransformer similarity are useful, they may not suffice in extractive tasks that require exact span matches. Conversely, exact match metrics may overly penalize semantically equivalent but paraphrased answers.

\subsection{User Evaluation Limitations}
Due to recruitment barriers, real physicians were not included in early testing. As a workaround, AI personas simulated realistic usage. However, this method’s reliability in mirroring true clinician needs remains uncertain. Further, prompt engineering for AI evaluation warrants standardization.

\section{Conclusion and Future Work}
This paper introduced a physician-centered QA system designed for unstructured clinical notes. The system supports text and voice queries, extractive highlighting, and adaptive question suggestion. Evaluations demonstrated encouraging performance across models and highlighted the promise of AI persona-based usability testing.

Future work will focus on validating AI persona evaluations through studies with real clinicians and enhancing model transparency by developing more robust explanation mechanisms. Additionally, we plan to explore interactive features tailored to the practical needs of physicians, aiming to improve clinical adoption and trust.

\bibliographystyle{IEEEtran}

\begin{thebibliography}{99}

\bibitem{b5} Jiang, E. (2024). Clinical Question-Answering over Distributed EHR Data. \textit{Ph.D. Thesis, MIT}.

\bibitem{b13} Bardhan, J., Roberts, K., \& Wang, D. Z. (2024). Question Answering for Electronic Health Records: Scoping Review of Datasets and Models. \textit{Journal of Medical Internet Research}, 26, e53636.

\bibitem{b4} Soni, S., Datta, S., \& Roberts, K. (2023). quEHRy: a question answering system to query electronic health records. \textit{Journal of the American Medical Informatics Association}, 30(6), 1091--1102.

\bibitem{b2} Bardhan, J., Colas, A., Roberts, K., \& Wang, D. Z. (2022). Drugehrqa: A question answering dataset on structured and unstructured electronic health records for medicine related queries. \textit{arXiv preprint arXiv:2205.01290}.

\bibitem{b3} Elgedawy, R., Danciu, I., Mahbub, M., \& Srinivasan, S. (2024). Dynamic Q\&A of Clinical Documents with Large Language Models. \textit{arXiv preprint arXiv:2401.10733}.

\bibitem{b6} Mahbub, M., Goethert, I., Danciu, I., Knight, K., Srinivasan, S., Tamang, S., Rozenberg-Ben-Dror, K., Solares, H., Martins, S., \& Trafton, J. (2024). Question-answering system extracts information on injection drug use from clinical notes. \textit{Communications Medicine}, 4(1), 61.

\bibitem{b7} Yu, H., Lee, M., Kaufman, D., Ely, J., Osheroff, J. A., Hripcsak, G., \& Cimino, J. (2007). Development, implementation, and a cognitive evaluation of a definitional question answering system for physicians. \textit{Journal of Biomedical Informatics}, 40(3), 236--251.

\bibitem{b9} Cao, Y., Liu, F., Simpson, P., Antieau, L., Bennett, A., Cimino, J. J., Ely, J., \& Yu, H. (2011). AskHERMES: An online question answering system for complex clinical questions. \textit{Journal of Biomedical Informatics}, 44(2), 277--288.

\bibitem{b10} Sarrouti, M., \& El Alaoui, S. O. (2020). SemBioNLQA: A semantic biomedical question answering system for retrieving exact and ideal answers to natural language questions. \textit{Artificial Intelligence in Medicine}, 102, 101767.

\bibitem{b11} Cai, P., Yao, Z., Liu, F., Wang, D., Reilly, M., Zhou, H., Li, L., Cao, Y., Kapoor, A., \& Bajracharya, A. (2023). Paniniqa: Enhancing patient education through interactive question answering. \textit{Transactions of the Association for Computational Linguistics}, 11, 1518--1536.

\bibitem{b12} Jimenez, E., \& Wu, H. (2024). emrQA-msquad: A Medical Dataset Structured with the SQuAD V2.0 Framework, Enriched with emrQA Medical Information. \textit{arXiv preprint arXiv:2404.12050}.

\bibitem{b1} Vong, W.-T., \& Then, P. H. H. (2015). Information seeking features of a PICO-based medical question-answering system. In \textit{2015 9th International Conference on IT in Asia (CITA)} (pp. 1--7). IEEE.

\bibitem{b16} IQA-EVAL: Automatic Evaluation of Human-Model Interactive Question Answering. 2023.

\bibitem{b17} OpenAI. (2025). OpenAI o3-mini. [Online]. Available: \url{https://openai.com/index/openai-o3-mini/}

\end{thebibliography}

\end{document}